# Effects of the Sun's time-retarded gravitational field on the orbital motions of Mercury and Halley's Comet

J. C. Hafele[1]

___

The neoclassical causal version for Newton's acausal gravitational theory explains exactly the anomalous speed-changes observed for six Earth flybys and an anomalous lunar orbital speed-change (arXiv:1105.3857v10). This article estimates the effects of the neoclassical causal theory on the orbital motions of two objects revolving around the Sun, Mercury and Halley's Comet. The change in the period for Mercury is predicted to be about +1.86 ms per year, and the predicted change in the angle for perihelion is -0.032 arc seconds per century, for which the magnitude is negligible compared with the relativistic advance of +43 arc seconds per century. The period for Halley's Comet, 75.3 years, is predicted to decrease by about 10 minutes. Therefore, the neoclassical causal theory does not conflict with general relativity theory, and it is not the cause for a delay of one or two weeks in the return time of Halley's Comet.
___

## 1. Introduction

The theory for the time-retarded gravitational field of a large spinning sphere has been shown to explain exactly the anomalous speed-changes observed by NASA for six Earth flybys, and it also explains exactly an anomalous change in the Moon's orbital speed.[1] The objective of this article is to estimate the effects of the Sun's time-retarded transverse gravitational field on the orbital motions of Mercury and Halley's Comet. The Sun's time-retarded gravitational field is derived in Section 2.

Mercury is the innermost and smallest planet in the solar system and has the largest eccentricity.[2] The observed precession of the perihelion is 5600 arc seconds per century, but the advance caused by the other planets calculated by using newtonian theory is 5557 arc seconds per century. The excess of 43 arc seconds per century is explained exactly by Einstein's general relativity theory. The change in the angle for perihelion caused by the neoclassical causal theory will be shown in Section 3 to be only -0.032 arc seconds per century. Therefore, there is no conflict between the neoclassical causal theory and general relativity theory.

Halley's Comet, the most famous of the "short-period" comets, has a period of 75.3 years.[3] There is a puzzling observed delay of one or two weeks in the return time.[4] The effect on the period caused by the neoclassical causal theory will be shown in Section 4 to decrease the period by about 10 minutes. Therefore, the delay predicted by the neoclassical causal theory is far too small to explain the observed delay of one or two weeks.

___
1. Retired PhD Professor of Physics and Math. Email: cahafele@bresnan.net.

## 2. Solar time-retarded gravitational field, speed change, and perihelion change for orbiting objects

Let the Sun's time-retarded transverse gravitational field be designated by $\mathbf{g}_{trt}$. The formula for the signed magnitude is [1]

$$g_{trt}(\theta) = -G \frac{I_S}{r_S^4} \frac{v_{Eq}}{c} \frac{\Omega_\phi(\theta) - \Omega_S}{\Omega_S} \cos^2(\lambda(\theta)) PS(r(\theta)) \; , \tag{2.1}$$

where G is the gravity constant, c is the vacuum speed of light, $I_S$ is the Sun's moment of inertia, $r_S$ is the Sun's radius, $\Omega_S$ is the Sun's spin angular speed, $v_{Eq}=r_S\Omega_S$ is the equatorial rotational surface speed, θ is the coordinate angle for the orbiting object in the plane of the orbit, $\Omega_\theta \equiv d\theta/dt$ is the orbital angular speed, $\Omega_\phi$ is the azimuthal φ-component of $\Omega_\theta$, λ is the heliocentric latitude of the orbiting object, r is the heliocentric radial distance of the orbiting object, and PS(r) is an inverse-cube power series representation for a triple integral over the Sun's volume. Numerical values for the Sun's parameters from Appendix A are

$$r_S = 6.955 \times 10^8 \text{ m} \; ,$$
$$\Omega_S = 2.911 \times 10^{-6} \text{ rad/s} \; ,$$
$$v_{Eq} = 2.025 \times 10^3 \text{ m/s} \; ,$$
$$I_S = 3.367 \times 10^{46} \text{ kg} \cdot \text{m}^2 \; . \tag{2.2}$$

The power series PS(r) is defined by

$$PS(r) = \left(\frac{r_S}{r}\right)^3 \left(C_0 + C_2 \left(\frac{r_S}{r}\right)^2 + C_4 \left(\frac{r_S}{r}\right)^4 + C_6 \left(\frac{r_S}{r}\right)^6\right) \; . \tag{2.3}$$

Numerical values for the coefficients from Appendix A are

$$C_0 = 0.500000 \; , \quad C_2 = 0.017498 \; ,$$
$$C_4 = 0.001376 \; , \quad C_6 = 0.000173 \; . \tag{2.4}$$

The formula for the "induction-like" field, $\mathbf{F}_\lambda$, is [1]

$$F_\lambda(\theta) = \frac{v_{Eq}}{v_k} \frac{r_E}{r(\theta)} \int_0^\theta \frac{r(\theta)}{r_E} \frac{\Omega_\theta(\theta)}{\Omega_E} \frac{1}{r_E} \frac{dr}{d\theta} \frac{dg_{trt}}{d\theta} d\theta \; , \tag{2.5}$$

where $v_k$ is an adjustable parameter called the "induction speed". According to ref. [1], $v_k$ probably lies between $2v_{Eq}$ and $8v_{Eq}$. The central value, $v_k=5v_{Eq}$, will be used herein.

The formulas for $r(\theta)$ and its derivative are

$$r(\theta) = \frac{r_p(1+\varepsilon)}{1+\varepsilon\cos(\theta)} \; , \quad \frac{dr}{d\theta} = \frac{r(\theta)^2 \varepsilon}{r_p(1+\varepsilon)} \sin(\theta) \; , \tag{2.6}$$

where ε is the eccentricity and $r_p$ is the radial distance at perihelion.

The calculated speed-change, $\delta v_{trt}$, is given by

$$\delta v_{trt} = \delta v_{in} + \delta v_{out} = \delta v(-\pi) + \delta v(+\pi) \, , \quad (2.7)$$

where

$$\delta v(\theta) = \frac{v_{in}}{2} \int_0^\theta \frac{r_\lambda F_\lambda}{v_{in}^2} \frac{d\lambda}{d\theta} \, d\theta \, . \quad (2.8)$$

Let a and b be the semimajor and semiminor axes for an elliptical orbit of eccentricity $\varepsilon$. Kepler's laws give the angular speed $\Omega_\theta$ and the period P in terms of a and b.[5]

$$a = \frac{1}{2}(r_a + r_p) \, , \qquad b = a(1 - \varepsilon^2)^{1/2} \, ,$$

$$P = \frac{2\pi a^{3/2}}{(GM_S)^{1/2}} \, , \qquad \Omega_\theta(\theta) = \frac{2\pi}{P} \frac{ab}{r(\theta)^2} \, , \quad (2.9)$$

where $r_a$ and $r_p$ are the heliocentric radial distance to the field-point at aphelion and perihelion.

Let $v_{co}$ be the orbital speed and let $\Omega_{co}$ be the orbital angular speed for a circular orbit of radius a and period P. The formulas for $v_{co}$ and $\Omega_{co}$ can be displayed by rearranging the formula for P (cf. Eq. (2.9)).

$$P = \frac{2\pi a}{(GM_S)^{1/2}} a^{1/2} = \frac{2\pi a}{(GM_S/a)^{1/2}} = \frac{2\pi a}{v_{co}} = \frac{2\pi}{\Omega_{co}} \, ,$$

$$v_{co}^2 = \frac{GM_S}{a} \, ,$$

$$\Omega_{co} = \frac{v_{co}}{a} \, . \quad (2.10)$$

Let $\delta v_{co} \ll v_{co}$ be a small change in $v_{co}$, let $\delta a \ll a$ be a corresponding change in a, let $\delta\Omega_{co} \ll \Omega_{co}$ be a corresponding change in $\Omega_{co}$, and let $\delta P \ll P$ be a corresponding change in P.

$$\left(1 + \frac{\delta v_{co}}{v_{co}}\right)^2 \cong 1 + 2\frac{\delta v_{co}}{v_{co}} = \frac{1}{(1 + \delta a/a)} \cong 1 - \frac{\delta a}{a} \, ,$$

$$\frac{\delta a}{a} \cong -2\frac{\delta v_{co}}{v_{co}} \, ,$$

$$\frac{\delta\Omega_{co}}{\Omega_{co}} \cong \frac{\delta v_{co}}{v_{co}} - \frac{\delta a}{a} = 3\frac{\delta v_{co}}{v_{co}} \, ,$$

$$\frac{\delta P}{P} \cong -\frac{\delta\Omega_{co}}{\Omega_{co}} = -3\frac{\delta v_{co}}{v_{co}} \, . \quad (2.11)$$

Let $\delta\theta_p \ll 2\pi$ be a small change in the value for $\theta$ at perihelion. Then

$$\frac{\delta\theta_p}{2\pi} \cong \frac{\delta\Omega_{co}}{\Omega_{co}} = 3\frac{\delta v_{co}}{v_{co}} \, . \quad (2.12)$$

These are the formulas that will be used to estimate the change in the period and the change in the angle for perihelion.

## 3. Change in Mercury's period and perihelion angle

Orbital parameters and the observed period for Mercury from Appendix B are

$$r_a = 69816900 \times 10^3 \text{ m} , \qquad r_p = 46001200 \times 10^3 \text{ m} ,$$
$$\varepsilon = 0.205630 , \qquad P_M = 87.969 \text{ days} = 7.6005 \times 10^6 \text{ s} ,$$
$$\alpha_{Eq} = 3.38° , \qquad \lambda_p = 3.38° . \tag{3.1}$$

Calculated values for the semimajor and semiminor axes and for the period are

$$a = \frac{1}{2}(r_a + r_p) = 5.791 \times 10^{10} \text{ m} ,$$
$$b = a(1 - \varepsilon^2)^{1/2} = 5.667 \times 10^{10} \text{ m} ,$$
$$P = 7.5998 \times 10^6 \text{ s} = 1.01 P_M . \tag{3.2}$$

Equations (2.6) and (2.9) give the orbital angular speed.

$$r(\theta) = \frac{r_p(1 + \varepsilon)}{1 + \varepsilon \cos(\theta)} ,$$
$$\Omega_\theta(\theta) = \frac{2\pi}{P} \frac{ab}{r(\theta)^2} . \tag{3.3}$$

Numerical values for the orbital angular speed and orbital speed at aphelion are

$$\Omega_a = \Omega_\theta(-\pi) = 5.559 \times 10^{-7} \text{ rad/s} ,$$
$$v_a = r_a \Omega_a = 3.881 \times 10^4 \text{ m/s} . \tag{3.4}$$

Values for the orbital speed and the orbital angular speed for the equivalent circular orbit are (*cf.* Eq. (2.10))

$$v_{co} = \left(\frac{GM_S}{a}\right)^{1/2} = 4.788 \times 10^4 \text{ m/s} ,$$
$$\Omega_{co} = \frac{v_{co}}{a} = 8.268 \times 10^{-7} \text{ rad/s} . \tag{3.5}$$

Mercury's heliocentric latitude is

$$\lambda(\theta) = \tan^{-1}\left(\tan(\alpha_{Eq}) \cos(\theta)\right) , \tag{3.6}$$

and the $\phi$-component of $\Omega_\theta$ is

$$\Omega_\phi(\theta) = \Omega_\theta(\theta) \cos(\alpha_{Eq}) . \tag{3.7}$$

The $\lambda$-component of r is

$$r_\lambda(\theta) = r(\theta) \cos(\theta) . \tag{3.8}$$

The formula for the Sun's time-retarded transverse field at Mercury's orbit is (*cf.* Eq. (2.1))

$$g_{trt}(\theta) = -G \frac{I_S}{r_S^4} \frac{v_{Eq}}{c} \left( \frac{\Omega_\phi(\theta) - \Omega_S}{\Omega_S} \right) \cos^2(\lambda(\theta)) PS(r(\theta)) \quad . \tag{3.9}$$

The formula for the induction-like field is

$$F_\lambda(\theta) = \frac{v_{Eq}}{v_k} \frac{r_S}{r(\theta)} \int_0^\theta \frac{r(\theta)}{r_S} \frac{\Omega_\theta(\theta)}{\Omega_S} \frac{1}{r_S} \frac{dr}{d\theta} \frac{dg_{trt}}{d\theta} d\theta \quad . \tag{3.10}$$

Choose $v_k = 5 v_{Eq}$ (*cf.* Eq. (2.5)).

By numerical integration, the speed-change becomes (*cf.* Eq. (2.7))

$$\delta v_{trt} = -9.428 \times 10^{-7} \text{ m/s per revolution} \quad . \tag{3.11}$$

Let $\theta_p$ be the value for $\theta$ at perihelion, let $\delta\theta_p$ be the change in $\theta_p$ per revolution. Set $\delta v_{co} = \delta v_{trt}$. Then

$$\frac{\delta\theta_p}{2\pi} = \frac{\delta\Omega_{co}}{\Omega_{co}} = 3 \frac{\delta v_{co}}{v_{co}} = 3 \frac{\delta v_{trt}}{v_{co}} \quad ,$$

$$\delta\theta_p = 6\pi \frac{\delta v_{trt}}{v_{co}} = -3.712 \times 10^{-10} \text{ rad per revolution} \quad . \tag{3.12}$$

Let $N_{rev}$ be the number of Mercury's revolutions in one year, and let $\Delta\theta_p$ be the accumulated angular change during one year.

$$N_{rev} = \frac{365.25}{87.969} = 4.125 \quad ,$$

$$\Delta\theta_p = N_{rev} \delta\theta_p = -1.541 \times 10^{-9} \text{ rad per year} \quad ,$$

$$\Delta\theta_p \times \frac{180}{\pi} \times 60 \times 60 \times 100 = -0.032 \text{ arc seconds per century} \quad . \tag{3.13}$$

Thus we find that the magnitude for the change in the angle for perihelion is less than 0.04 arc seconds per century, which is totally negligible compared with the relativistic change of 43 arc seconds per century.

The change in the period is (*cf.* Eq. (2.11))

$$\delta P = -3P \frac{\delta v_{trt}}{v_{co}} N_{rev} = 1.86 \text{ ms per year} \quad . \tag{3.14}$$

During the next 1000 years, the period is predicted to increase by about 2 seconds.

### 4. Change in the period for Halley's Comet

Numerical values for the orbital parameters and observed period for Halley's Comet from Appendix C are

$$r_a = 35.1 \text{AU} = 5.251 \times 10^{12} \text{ m} \quad , \quad r_p = 0.586 \text{AU} = 8.766 \times 10^{10} \text{ m} \quad ,$$

$$\varepsilon = 0.967 \quad , \quad P_{HC} = 75.3 \text{ years} = 2.376 \times 10^9 \text{ s} \quad ,$$

$$\alpha_{Eq} = 162.3° \quad\quad \lambda_p = 180° - 162.3° = 17.7° \quad . \tag{4.1}$$

Calculated values for the semimajor and semiminor axes and for the period are

$$a = \frac{1}{2}(r_a + r_p) = 2.669 \times 10^{12} \text{ m},$$
$$b = a(1 - \varepsilon^2)^{1/2} = 6.785 \times 10^{11} \text{ m},$$
$$P = 2.378 \times 10^9 \text{ s} = 1.001 P_{HC}. \tag{4.2}$$

Equations (2.6) and (2.9) give the orbital angular speed.

$$r(\theta) = \frac{r_p(1+\varepsilon)}{1+\varepsilon \cos(\theta)},$$
$$\Omega_\theta(\theta) = \frac{2\pi}{P}\frac{ab}{r(\theta)^2}. \tag{4.3}$$

Numerical values for the orbital angular speed and orbital speed at aphelion are

$$\Omega_a = \Omega_\theta(-\pi) = 1.735 \times 10^{-10} \text{ rad/s},$$
$$v_a = r_a \Omega_a = 9.112 \times 10^2 \text{ m/s}. \tag{4.4}$$

Values for the orbital speed and the orbital angular speed for the equivalent circular orbit are (*cf.* Eq. (2.10))

$$v_{co} = \left(\frac{GM_S}{a}\right)^{1/2} = 7.052 \times 10^3 \text{ m/s},$$
$$\Omega_{co} = \frac{v_{co}}{a} = 2.642 \times 10^{-9} \text{ rad/s}. \tag{4.5}$$

The $\phi$-component of $\Omega_\theta(\theta)$ reduces to (the sign is negative because the orbital motion is retrograde)

$$\Omega_\phi(\theta) = -\Omega_\theta(\theta)\cos(\lambda_p). \tag{4.6}$$

The heliocentric latitude, $\lambda(\theta)$, is

$$\lambda(\theta) = \tan^{-1}\left(\tan(\lambda_p)\cos(\theta)\right). \tag{4.7}$$

The formula for the Sun's time-retarded transverse field at the orbit for Halley's Comet is (*cf.* Eq. (2.1))

$$g_{trt}(\theta) = -G\frac{I_S}{r_S^4}\frac{v_{Eq}}{c}\left(\frac{\Omega_\phi(\theta) - \Omega_S}{\Omega_S}\right)\cos^2(\lambda(\theta))PS(r(\theta)). \tag{4.8}$$

The formula for the induction-like field is

$$F_\lambda(\theta) = \frac{v_{Eq}}{v_k}\frac{r_S}{r(\theta)}\int_0^\theta \frac{r(\theta)}{r_S}\frac{\Omega_\theta(\theta)}{\Omega_S}\frac{1}{r_S}\frac{dr}{d\theta}\frac{dg_{trt}}{d\theta}d\theta. \tag{4.9}$$

Choose the same value as was used for Mercury, $v_k = 5v_{Eq}$.

The λ-component of r(θ) reduces to

$$r_\lambda(\theta) = r(\theta) \cos(\theta) \ . \tag{4.10}$$

The formulas for $\delta v(\theta)$ and $\delta v_{trt}$, and the numerical value for $\delta v_{trt}$ by numerical integration, are

$$\delta v(\theta) = \frac{v_a}{2} \int_0^\theta \frac{r_\lambda(\theta) F_\lambda(\theta)}{v_a^2} \frac{d\lambda}{d\theta} d\theta \ ,$$

$$\delta v_{trt} = \delta v(-\pi) + \delta v(+\pi) = +5.969 \times 10^{-4} \text{ m/s per period} \ . \tag{4.11}$$

The formula for the change in the period is (*cf.* Eq. (2.11))

$$\delta P = -3P \frac{\delta v_{trt}}{v_{co}} = -604 \text{ s} = -10 \text{ minutes} \ . \tag{4.12}$$

Thus we find that the predicted decrease in the period of 10 minutes is way too small to account for a one or two week delay in the return of Halley's Comet.

## 5. Conclusions and recommendations

It has been shown herein that the Sun's time-retarded transverse gravitational field causes a change in the angle for the perihelion of Mercury of only about -0.032 arc seconds per century, and a decrease in the period for Halley's Comet of only about 10 minutes. Such small changes are undetectable. Therefore, it is recommended that we look elsewhere for confirmation of time-retardation effects.

## Acknowledgements

I thank Patrick L. Ivers for reviewing the original manuscript and suggesting improvements.

## Appendix A: Numerical values mostly for the Sun

The following numerical values are available in handbooks.

$G = 6.6732 \times 10^{-11} \frac{m^3}{kg \cdot s^2}$  Gravity constant

$c = 2.997925 \times 10^8$ m/s  Vacuum speed of light

$AU = 149\,597\,870.7 \times 10^3$ m  Astronomical Unit

The following numerical values can be found in the Wikipedia article "Sun".[6]

$r_S = 6.955 \times 10^8$ m  Equatorial radius

$M_S = 1.9891 \times 10^{30}$ kg  Mass

$V_S = 1.412 \times 10^{27}$ m$^3$  Volume

$\bar{\rho} = 1.408 \times 10^3$ kg/m$^3$  Mean mass-density

$\rho_{ctr} = 1.622 \times 10^5$ kg/m$^3$  Mass-density at the center

$\rho_{photo} = 2 \times 10^{-4}$ kg/m$^3$  Mass-density at the photosphere

$P_S = 25.05$ days $= 2.158 \times 10^6$ s  Equatorial rotational period

$$\alpha_S = 7.25°  \quad \text{Obliquity to the ecliptic}$$

The following values were computed from the above values.

$$\Omega_S = 2\pi/P_S = 2.911 \times 10^{-6} \text{ rad/s} \quad \text{Equatorial angular speed}$$

$$v_{Eq} = r_S \Omega_S = 2.025 \times 10^3 \text{ m/s} \quad \text{Rotational equatorial surface speed}$$

Let $I_S$ be the Sun's moment of inertia. The following exponential function for $\rho(r)$ provides a reasonably valid approximation for the Sun's radial mass-density distribution.

$$\rho(r) = \text{if}\left(r \leq r_S, \rho_{ctr} \exp\left(-\left(\frac{r}{r_{core}}\right)^2\right), 0\right) . \tag{B1}$$

The calculated mass is given by

$$M = \frac{4\pi}{3} \bar{\rho}_S r_S^3 \int_0^{r_S} \frac{\rho_S(r)}{\bar{\rho}_S} \frac{r^2}{r_S^3} \, dr . \tag{B2}$$

The following value for $r_{core}$,

$$r_{core} = 0.18707 , \tag{B3}$$

gives

$$(M - M_S)/M_S = 5 \times 10^{-5} , \tag{B4}$$

and

$$\rho_S(0.847 r_S) = 2.03 \times 10^{-4} \text{ kg/m}^3 = \rho_{photo} . \tag{B5}$$

The value for the spin moment of inertia using Eq. (B1) and numerical integration is

$$I_S = \frac{8\pi}{3} \bar{\rho} r_S^5 \int_0^{r_S} \frac{\rho(r)}{\bar{\rho}} \frac{r^4}{r_S^5} \, dr = 3.367 \times 10^{46} \text{ kg} \cdot \text{m}^2 . \tag{B6}$$

A four-term power series provides an excellent representation for the triple integral over the Sun's volume (*cf.* Eq. (1.1)).

$$\left(\frac{I_S}{\bar{\rho} r_S^5}\right) PS(r) = \left(\frac{I_S}{\bar{\rho} r_S^5}\right)\left(\frac{r_S}{r}\right)^3 \left(C_0 + C_2 \left(\frac{r_S}{r}\right)^2 + C_4 \left(\frac{r_S}{r}\right)^4 + C_6 \left(\frac{r_S}{r}\right)^6\right)$$

$$= \left(\frac{r_S}{r}\right)^3 \int_0^{r_S} \left(\int_{-\pi/2}^{\pi/2} \left(\int_{-\pi}^{\pi} \frac{\sin(\phi')}{(1+x)^2} d\phi'\right) \cos^3(\lambda') d\lambda'\right) \frac{\rho(r)}{\bar{\rho}} \frac{r^4}{r_S^5} \, dr , \tag{B7}$$

where the variable x is defined by

$$x \equiv \left(\frac{r}{r}\right)^2 - 2 \frac{r}{r} \cos(\lambda') \cos(\phi') . \tag{B8}$$

By using a least-squares fitting routine, the following numerical values were found to give an excellent fit of PS(r) to the volume integral.

$$C_0 = 0.500000, \quad C_2 = 0.017498,$$
$$C_4 = 0.001376, \quad C_6 = 0.000173. \tag{B9}$$

## Appendix B: Numerical values for the planet Mercury

The following numerical values can be found in the Wikipedia article "Mercury(planet)".[2]

| | | |
|---|---|---|
| $r_a = 69816900 \times 10^3$ m | | Aphelion |
| $r_p = 46001200 \times 10^3$ m | | Perihelion |
| $\varepsilon = 0.205630$ | | Eccentricity |
| $m = 3.3022 \times 10^{23}$ kg | | Mass |
| $P_M = 87.969$ days $= 7.6005 \times 10^6$ s | | Sidereal orbital period |
| $\alpha_{Eq} = 3.38°$ | | Inclination to Sun's equator |
| $\lambda_p = 3.38°$ | | Heliocentric latitude at perihelion |
| $\delta\theta_p = 43$ arc seconds per century | | Relativistic advance of perihelion |

## Appendix C: Numerical values for Halley's Comet

The following numerical values can be found in the Wikipedia article "Halley's Comet".[3]

| | | |
|---|---|---|
| $r_a = 35.1$ AU | | Aphelion |
| $r_p = 0.586$ AU | | Perihelion |
| $\varepsilon = 0.967$ | | Eccentricity |
| $m = 2.2 \times 10^{14}$ kg | | Mass |
| $P_{obs} = 75.3$ years $= 2.376 \times 10^9$ s | | Average observed orbital period |
| $\alpha_{HC} = 162.3°$ | | Inclination to ecliptic |
| $\lambda_p = 180° - 162.3° = 17.7°$ | | Heliocentric latitude at perihelion |